# Distributions of number of sexual partnerships have power law decaying tails and finite variance

Fredrik Liljeros[*,†], Christofer R. Edling[†], H. Eugene Stanley[‡], Y. Åberg[†], and Luís A. Nunes Amaral[**]

*Department of Medical Epidemiology and Biostatistics, Karolinska Institutet, SE-171 77 Stockholm, Sweden <liljeros@sociology.su.se>
† Department of Sociology, Stockholm University, S-106 91 Stockholm, Sweden.
‡ Centre for Polymer Studies, Department of Physics, Boston University, Boston, MA 02215, USA.
** Department of Chemical Engineering, Northwestern University, Evanston, IL 60208, USA.

In a recent working paper [1], James Holland Jones and Mark Handcock re-analyze two of the four datasets comprising a database, first analyzed by us in Ref. 2, which records the number of distinct sexual partners for Swedish men and women. Specifically, Jones and Handcock analyze the probability distribution function $p(k)$, a function from which one obtains less reliable information than from the cumulative distribution,

$$P(\geq k) = \sum_{k' \geq k} p(k'),$$

analyzed by us [2]. Jones and Handcock propose that the tails of $p(k)$ decay with a power law exponent $r > r_c = 3$, larger than the threshold value above which the variance of the distribution is finite.

In the following, we argue that the claims of Jones and Handcock can be interpreted in a misleading fashion. The two datasets analyzed by Jones and Handcock pertain to the number of *distinct* sexual partners during the previous year. This fact renders any discussion about the existence of epidemic thresholds speculative, since other factors besides the distribution of number of distinct sexual partners in the previous year control the existence of an epidemic threshold for infinitely large networks [3-7]. In finite networks there will always be a finite threshold, because in a finite network the variance of the distribution of number of partners must be finite.

Consider now Jones and Handcock's statement that $r > r_c = 3$. This is not a new result. The estimates by Jones and Handcock in fact agree with the estimates already reported by us in Ref. 2, which demonstrated that the tails of the distributions of the number of distinct sexual partners decay asymptotically as power laws. A similar power law tail has also been reported for clinical data on homosexual men [8]. The reason why the result of Jones and Handcock at first glance might appear to disagree with that of Ref. 2 arises from the fact that we study the cumulative distribution, $P(\geq k)$ — the fraction of events larger or equal to $k$ — while Jones and Handcock study the probability distribution function $p(k)$. Since $P(k)$ is a summation of $p(k)$, the exponent $r$ is related to the exponent $a$ by the equation $r = a + 1$. Hence the values of $a$ that are somewhat above 2 reported in Ref. 2 do not differ from the values of $r$ that are somewhat above 3 reported in Ref 1. The above statements are supported by Table 1, which presents estimates of the parameters for the power law tails using maximum likelihood estimation (MLE) and bootstrapped confidence intervals.

Another difference between the analyses of Refs. 1 and 2 is that Jones and Handcock assume that the data are best described by a Yule distribution [9]. However, the Yule distribution does not display a power law for small $k$ (Fig.1). Furthermore, the assumptions going into the derivation of the Yule distribution do not take into consideration that new contacts cannot be added to already existing partnerships, nor does the analysis consider the number of partners that individuals have at the beginning of the time interval analyzed. As social scientists are no longer forced to select mathematical models on the basis of analytical tractability at the cost of realism [10], it is surprising that Jones and Handock do not select from the more general models including preferential attachment published in the scientific literature [11].

**Table 1. Estimates of the tail exponent $r$ and confidence intervals for the distribution of distinct sexual partnerships for Swedish men and women.**

|  | No. of distinct partners during previous year | | | No. of distinct partners in lifetime | | |
|---|---|---|---|---|---|---|
|  | OLS Power law ($k_{men}>5$) ($k_{women}>4$) | MLE Power law ($k>0$) | MLE Yule ($k>0$) | OLS Power law ($20<k_{men}<400$) ($k_{women}>20$) | MLE Power law ($k>20$) | MLE Yule ($k>20$) |
| $r_{men}$ | 3.31 (3.11-3.51) | 2.73 (2.65-2.84) | 4.13 (4.0-4.7) | 2.16 (1.86-2.46) | 2.37 (2.26-2.50) | 2.40 (2.28-2.53) |
| $r_{women}$ | 3.54 (3.34-3.74) | 3.26 (3.10-3.41) | 6.23 (5.54-6.98) | 3.1 (2.8-3.4) | 3.125 (2.77-3.62) | 3.21 (2.85-3.68) |

Note: MLE (Maximum Likelihood Estimation), OLS (Ordinary Least Squares regression).

Jones and Handcock also choose to ignore two of the datasets analyzed in Ref. 2, those concerning the number of *lifetime* partners. Such a practice of limiting one's analysis to a subset of the data (viz., number of partners during a one year period) places unnecessary restrictions on understanding the structure of the network of sexual contacts and is particularly surprising in view of the duration of the infectiousness period of some sexually transmitted infections such as HIV. Moreover, the distribution of the number of lifetime partners covers a broader range of $k$, and the analysis in Ref. 2 provides striking support for the hypothesis that the distributions of number of distinct sexual partners decay as a power law. For these reasons, we re-analyze the datasets for the number of distinct lifetime sexual partners using the Yule distribution and find the same result as for a power law fit (Table 1, columns 4-6). This result does not come as a surprise since the Yule distribution closely resembles a power law in the interval analyzed.

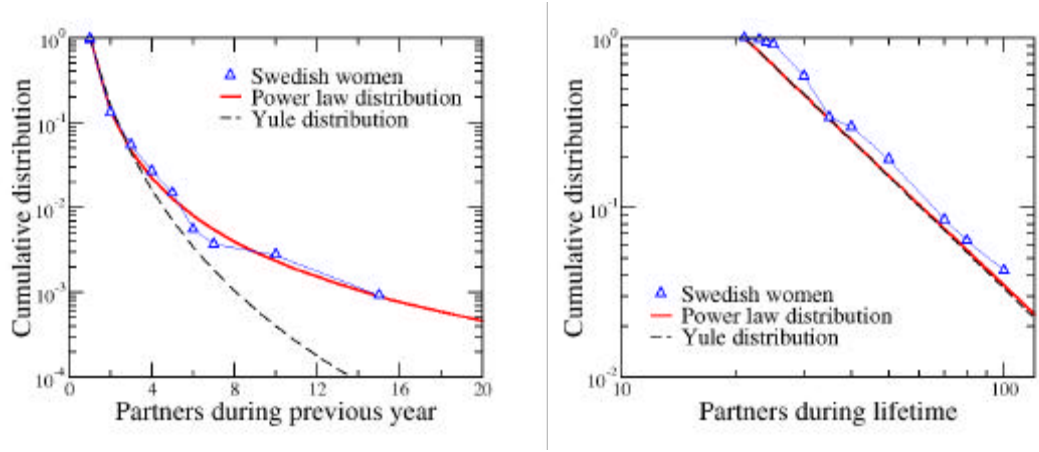

Fig 1. Comparison of the fitting of the data to both Yule and power law distributions. We show fits to the cumulative distribution of number of distinct sexual partners for Swedish women during the previous year (log-linear plot) and for lifetime (log-log plot). Note that for number of sexual partners during the previous year the agreement with a power law distribution is much better than with the Yule distribution, while for the number of lifetime partners both distributions lead to nearly identical estimates.

Finally, Jones and Handcock argue that if the variance of the distribution of sexual partners is finite, interventions aimed at reducing disease transmissibility have the potential to eradicate sexually transmitted infections. As we stated earlier, the variance of these distributions must be finite since we are analyzing finite populations of individuals that are sexually active for finite periods of time [12]. However, Jones and Handcock's conclusion is somewhat premature since research convincingly demonstrates that contagious processes differ fundamentally between networks with power law decaying degree distributions and networks with fast decaying degree distributions [7,13]. Specifically, if a detectable epidemic threshold exists, it will certainly be very small as the variance of the distribution of number of sexual partners is much larger than the mean [13-15].